\documentstyle[aps,multicol,prb,epsf]{revtex}
\newcommand{\pp}{^{\prime \prime}}

\begin{document}
\title{Theory of the Spin Gap in High Temperature Superconductors}
\author{B. L. Altshuler}
\address{NEC Research Institute, 4 Independence Way, Princeton, NJ 08540 \\
Physics Department, MIT, Cambridge, MA 02139}

\author{L. B. Ioffe}
\address{Physics Department, Rutgers University, Piscataway, NJ 08855 \\
and Landau Institute for Theoretical Physics, Moscow}

\author{A. J. Millis}
\address{AT\&T Bell Laboratories, Murray Hill, NJ 07974}
\maketitle
\begin{abstract}
We analyze pairing in two dimensional spin liquids.
We argue that interplane pairing enhanced by magnetic correlations is the most
plausible explanation of the spin gap phenomenon observed in underdoped
cuprates.
The details of the pairing theory depend on the in-plane antiferromagnetic
correlations.
We consider two models: $2p_F$ correlations induced by a strong gauge field
interaction and undamped spin waves.
We estimate the pairing temperature $T_s$ and the angular dependence of the
gap function and discuss physical consequences.
\end{abstract}
\pacs{74.20.Mn, 76.60-k, 75.10.Lp, 74.72.-h}

\begin{multicols}{2}

\section{INTRODUCTION}
Many high $T_c$ materials exhibit anomalous temperature dependence of the
bulk magnetic susceptibility, $\chi$, in a range of temperatures above $T_c$,
suggesting that a spin pseudogap opens above the superconducting transition
temperature \cite{Rice89}.
To define more precisely what we mean by spin pseudogap consider the
susceptibility data for $YBa_2Cu_4O_8$ shown as the curve labeled ``248'' in
Fig. 1.
(The data points shown in Fig. 1 were obtained from the susceptibility
reported in Ref. \cite{248chi,214chi} as discussed in Ref. \cite{Millis93a})
There are clearly two regimes, separated by a scale $T_s \approx 200$K.
For $T>T_s$, $\chi \cong A+BT$ with $A$, $B>0$.
For $T<T_s$, $\chi(T)$ drops more rapidly; indeed a straight line fit to
$\chi(T)$ for $T_c<T<T_s$ ($T_c = 80$K is the superconducting
transition temperature for this compound) would yield a {\it negative}
$\chi$ at $T=0$.
A negative $\chi(0)$ is impossible; therefore the negative extrapolation
implies that even for $T>T_c$ (but $T<T_s$) there is a gap for spin
excitations.
The origin of this ``spin'' gap is the main focus of this paper.

The properties of high temperature superconducting materials are anomalous,
and many different theories have been proposed to describe them.
However, no general consensus has emerged.
The origin of spin gap has been discussed by many authors
\cite{Rice89,Altshuler92,Sachdev92a,Tanamoto92,Randeira92,Millis93a,Sokol93,Ioffe94,Millis95a}
but these treatments are not completely satisfactory because, as we shall
argue below, they are based on models which do not agree with all available
data.
We believe that any theory of the spin gap should have the following
ingredients:
\begin{itemize}
\item [(i)]
At least some of the magnetic response is Pauli-like, i.e. it comes from a
particle-hole continuum of spin $1/2$ fermions.
\item [(ii)]
Formation of the spin gap involves pairing instability of these fermions.
\item [(iii)]
This pairing is {\it not} a superconducting pairing, i.e. does not produce a
Meissner effect or paraconductivity.
\end{itemize}
These assumptions imply that paired fermions are neutral and that any theory
of the spin gap must involve the phenomenon of spin-charge separation; we use
the gauge  theory formalism to describe this.

\begin{figure}
\vspace{-2.5cm}
\centerline{\epsfxsize=8cm \epsfbox{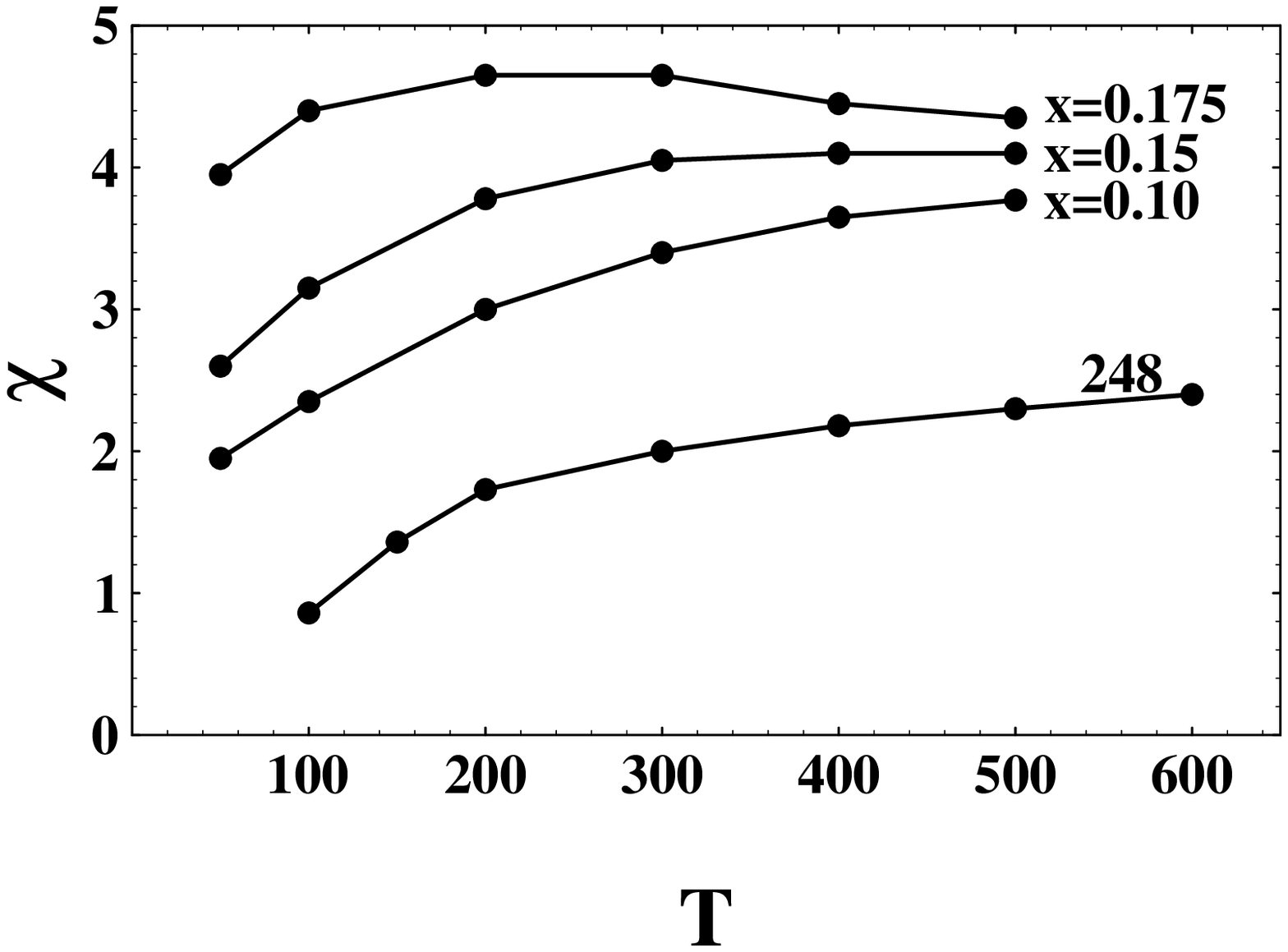}}
\vspace{-2.5cm}
{
FIG 1.
Susceptibility of several high $T_c$ materials: one-plane $La_{2-x}Sr_xCuO_4$
with varying doping $x$ which, as we shall argue, does not show spin gap
behavior and bilayer $YBa_2Cu_4O_8$ (248) which shows it.
Susceptibility is measured in the units of $\mu_B^2 \; eV$ per $Cu$ atom.
}
\end{figure}
\vspace{0.2in}

Points (i)-(iii) do not completely specify the model.
There are two additional issues.
\begin{itemize}
\item [(iv)]
It is widely believed that there are strong antiferromagnetic spin
fluctuations in high-$T_c$ materials.
In the body of the paper we summarize the evidence for the existence of
antiferromagnetic spin fluctuations, outline the different theories proposed
to  describe them, and give the implications for our calculations.
\item [(v)]
We believe that bilayer structure of $YBa_2Cu_4O_8$ is important for the
formation of the spin gap because it leads to interplane pairing.
Whether the spin gap exists in single-plane materials such as
$La_{2-x}Sr_xCuO_4$, in which interplane coupling is geometrically frustrated,
is controversial \cite{Millis93a,Sokol93}.
Certainly, the evidence for it is weaker (see Fig. 1).
Our theoretical results imply that spin gap effects are greatly enhanced in
bilayer or multilayer systems.
\end{itemize}

The outline of the paper is as follows.
In Section II, we review the relevant experimental data and show that it
implies points (i)-(v) above.
In Section III we formulate the theoretical model and identify two
different scenarios.
In Section IV and V we present the solutions for the different scenarios.
Section V contains a comparison of results to data and a conclusion.

\section{DATA}

\noindent
(i) {\it Particle-hole continuum}:
We begin with the evidence that at least some of the magnetic response in
spin-gap systems such as $YBa_2Cu_4O_8$ comes from a particle-hole
continuum of spin $1/2$ fermions.
A phenomenological argument is that there is little doubt that such a
continuum describes the magnetic properties of $YBa_2Cu_3O_7$.
The uniform susceptibility of this compound has magnitude slightly larger than
predicted by band theory and has very weak temperature dependence (as expected
for a Fermi liquid) \cite{chiO7}.
The oxygen NMR relaxation rate $1/ ^{17}T_1T$ is temperature independent
(as expected in a Fermi liquid) and has magnitude slightly larger than
predicted by the Korringa relation \cite{OT1TO7}.
The copper relaxation rate $1/ ^{63}T_1 T$ increases as the
temperature is decreased \cite{CuT1TO7}; this has been argued to be due to
antiferromagnetic correlations within a Fermi liquid state \cite{Millis90}.
Also, photoemission experiments \cite{Campuzano92} have observed a large Fermi
surface in optimally doped $YBa_2Cu_3O_7$.
The spin-gap compounds $YBa_2Cu_3O_{6.6}$ and $YBa_2Cu_4O_8$ are produced by
removing carriers from $YBa_2Cu_3O_7$; further, the physical properties seem
to vary smoothly with doping.
Therefore, it is natural to assume that the particle hole continuum is also
present in the spin-gap compounds.

An alternative point of view \cite{Sokol93} is that the important
spin excitations in underdoped compounds are weakly damped, gapped
antiferromagnetic spin waves, with dispersion  $\omega^2 = c^2k^2+ \Delta^2$.
However, theories in which the only magnetic excitations are spin waves with a
gap are inconsistent with the $\chi_s(T)$ data at {\em high temperatures}
($T>T_s$), because at temperatures greater than the value of
the zero-temperature gap $\Delta_0$, these theories predict \cite{Sachdev92b}
$\chi_s = A + BT$ with $A < 0$ if $\Delta_0 > 0$ and also predict for
$YBa_2Cu_4O_8$  a value for $B$ larger, by a factor of six, than is
observed \cite{Millis94}.
Another difficulty with the spin-wave-only model is that the kinematics of
spin waves implies that the oxygen relaxation rate drops very rapidly with
temperature.
Indeed the theoretical prediction is $1/[T_1 T \chi_s ] \sim T$ in contrast to
the experimental result $1/ (T_1T \chi_s ) \; \approx \; constant$.
\cite{Millis94}
Therefore there is little doubt that one must at least supplement the spin
wave model with a particle-hole continuum of fermions which controls at least
the small $q$ spin response \cite{Millis93b,Chubukov93}.
We note, however, that although the existence of such a particle-hole continuum
would increase the constant part, $A$, of the uniform susceptibility and would
add a constant contribution to the
oxygen relaxation rate, it would not affect the slope, $B$, of the
susceptibility  if the interaction between spin waves and particle-hole
continuum is weak.
There is as yet no model which describes all of the data.
In any event, to understand the spin gap observed in $\chi_s (T)$ one must
understand how to open a gap in the spectrum of the fermions making up the
particle-hole continuum.

\noindent
(ii) {\it Singlet pairing}: In the absence of charge density wave order
the only possibility for suppressing the spin response of fermions is to pair
them into singlets.
One example of such pairing is the BCS superconducting state.
The pairing scale $T_s \approx 200 \; K$ is much less than any microscopic
scale such as $J$ or the bandwidth, so the pairing must be understood as a low
energy instability of the system.

\noindent
(iii) {\it Superconductivity}:
The data strongly suggest that the spin gap is not due to incipient
superconducting pairing.
There are two arguments; one is that the spin gap scale $T_s$ in
$YBa_2Cu_4O_8$ is approximately $200$K, far above the maximum $T_c \approx 95$K
observed in any member of the YBCO family.
The second is that superconducting fluctuations normally have a dramatic
effect on the resistivity, which begins to drop as $T$ is decreased through
the scale at which the fluctuations begin and drops more and more rapidly
as $T_c$ is approached.
This behavior is not observed in $YBa_2Cu_3O_{6.6}$ and in $YBa_2Cu_4O_8$.
In these materials, there is some drop in $\rho$ as $T$ is decreased through
$T_s$, but then $\rho$ flattens out and depends only weakly on $T$
near $T_c$ \cite{rhoO6.6,rho248}.
The drop in $\rho$ for $T \sim T_s$ has been attributed \cite{rho248}, in our
view correctly, to changes in the inelastic scattering mechanism associated
with the onset of the spin gap.
It has been argued that negative $U$ Hubbard model, in which the important
physics is singlet pairing of conventional electrons describes the spin gap
phenomena \cite{Randeira92}.
In our view the strong evidence for repulsive interactions and against
paraconductivity renders this model irrelevant.

To summarize:
the data imply the existence of a particle-hole continuum of spin
excitations and require that these excitations be
paired into singlets in a way that does not
produce superconductivity.
This implies that ``spin-charge separation'' must occur.
There are many scenarios of spin-charge separation, all stemming from
Anderson's original proposal \cite{Anderson87}.
We shall adopt the gauge theory approach \cite{Ioffe89,Baskaran87}.
We also note that unlike the conventional superconducting pairing of
electrons, the pairing of chargeless fermions does not necessary imply the
breaking of any symmetry, and so may result in a crossover, not in a genuine
phase transition \cite{Altshuler92}.

\noindent
(iv) {\it Antiferromagnetism}:
A large literature has developed around the issue of antiferromagnetic
correlations in high $T_c$ materials.
Many controversies remain unresolved, but it is generally agreed that NMR
$T_1$ and $T_2$ experiments imply the existence of strong, temperature
dependent antiferromagnetic correlations.
Specifically, for $YBa_2Cu_4O_8$, the two $Cu$ relaxation rates, $1/T_1T$ and
$1/T_2$, increase roughly as  $1/T$ as $T$ is decreased (for $T >T_s$)
\cite{248T1,248T2}.
Generally, the NMR rates are given by
\begin{eqnarray}
\frac{1}{T_1T} &=& \lim_{\omega \rightarrow 0}
\sum_q F_q \frac{\chi\pp (q, \omega )}{\omega}
\label{T_1T} \\
\frac{1}{T_2} &=& \left [ \sum_q
[F_q \chi^{\prime} (q, 0) ]^2 \right ]^{1/2}
\label{T_2}
\end{eqnarray}
where $F_q$ is a form factor which is different for different nuclei.

The only tenable interpretation of the temperature dependence of the copper
relaxation rate is that both real and imaginary parts of $\chi({\bf q})$
diverge at a particular wavevector ${\bf Q}$, i.e. that there is an incipient
magnetic instability.
To prove this, suppose on the contrary that the $T$-dependence
of $1/ (T_1 T)$ came from a wide range of momenta, so  $\chi(q, \omega ) =
\phi (q) f(\omega, T)$, with $\phi (q)$ a temperature independent function.
For large frequencies, $\omega \gg T$, $f(\omega)$ should not depend on
temperature, so in particular the imaginary part $ f''( \omega ) = A_1
\omega^x$, whereas at small frequencies,
$\omega \ll T$,  $f \pp (\omega)$ should be proportional to frequency,
$f''( \omega )=A_2 \omega$.
The proportionality coefficient, $A_2$, can be estimated by matching the low
frequency formula and the high frequency formula at $\omega \approx T$,
yielding $A_2=A_1 T^x$.
The NMR $T_1$ data imply $x \approx 0$ which via the Kramers-Kronig
relation implies $\chi^{\prime} (q, \omega =0) \sim \ln T$.
This temperature dependence is too weak to account for the $T_2$ data, so the
hypothesis of a momentum-independent divergence of $\chi$ must be rejected.
It should be noted, however, that although neutron scattering experiments
detect antiferromagnetic fluctuations, neutron and NMR data are not
quantitatively consistent \cite{Walstedt94,Rossat-Mignon91}

The proper theoretical model for the antiferromagnetic fluctuations
is not clear.
The two principal proposals are that the dominant antiferromagnetic
excitations are weakly damped spin waves \cite{Sokol93,Sachdev92b}, or are
particle-hole pairs of an antiferromagnetically correlated fermion system
\cite{Millis90}.
The weakly damped spin wave picture applies to the magnetic insulating parent
compound and by continuity might be expected to apply to lightly doped but
non-ordered materials.
The particle-hole picture presumably applies to the optimally doped materials
(which have been shown by photoemission to have large (Luttinger) Fermi
surface) and by continuity might be expected to apply to somewhat underdoped
materials.
The crossover between these two regimes is a active area of research
\cite{Sachdev95} but has not been understood in detail.
We consider implications of both pictures for the pairing interaction.

The model of particle-hole pairs requires further discussion.
We argued above on the basis of resistivity data that the fermions must be
charge zero objects.
An additional argument against conventional charge $e$ Fermi liquid with
antiferromagnetic correlations is that the magnetic properties of conventional
fermi liquid are incompatible with the $T_1$, $T_2$ and $\chi_s$ data
\cite{248T1,248T2,248chi,Ioffe95}.
The only known model of spin $1/2$ charge $0$ fermionic excitations
(``spinons'') which does not break time reversal or parity symmetry is the
``spin liquid''.
In a spin liquid, the gauge interaction between fermions affects the magnetic
properties and has been shown to lead to a reasonable description of the spin
dynamics of high $T_c$ materials \cite{Altshuler95}
In the spin-wave picture a particle-hole continuum is also present; for the
reasons given above we must assume that the underlying fermionic excitations
are spinons.
However, in this case the spin waves dominate the large $q$ magnetic response.

\noindent
(v) {\it Bilayers}:
Our discussion so far has been focussed on members of the $Y-Ba$ family of
high $T_c$ materials.
We believe that the magnetic dynamics of members of other families of high
$T_c$ materials are similar in all respects except for the existence of a spin
gap {\em in a wide temperature range above $T_c$}.
It is possible that, as some authors have argued \cite{Sokol93,Barzykin95},
$La_{2-x}Sr_xCuO_4$ shows the beginning of spin gap behavior for $T \sim 60$K,
relatively near the superconducting $T_c$.
However, to our knowledge only underdoped members of the $Y-Ba$ family exhibit
spin gap behavior over a wide range of temperatures above $T_c$.

Consider first the $\chi_s$ for $La_{2-x}Sr_xCuO_4$ data presented in Fig. 1.
These are obtained from bulk susceptibility data by subtracting core and
van-Vleck susceptibilities given in \cite{Johnston91}.
Note that although there is a downturn in $\chi_s (T)$ at $T<100$K,
extrapolated value of $\lim_{T \rightarrow 0} \chi_s (T)$ is positive and
relatively large.
Although uncertainties in the value of the van-Vleck susceptibility may exist,
these are in our view  by no means large enough to produce a
$lim_{T \rightarrow 0} \chi_s (T) \leq 0$.
The oxygen \cite{Walstedt94} and copper \cite{Walstedt94,Imai} $T_1$
relaxation data similarly show no sign of an extra {\it downturn} at a
temperature $T>T_c$ (although the rate of increase of the copper $T_1^{-1}$
may slow for $T<80 \;K$).
Note that at $T>T_s \approx  T_c$, $La_{2-x}Sr_xCuO_4$ does exhibit  $\chi_s
\approx A+BT$ regime as well as other properties difficult to describe in
either a Fermi liquid or a purely spin wave picture.
We infer from this that the bilayer structure of the Y-Ba material is
important only for raising $T_s$ sufficiently far above $T_c$  that spin gap
effects are easily observable.

There is in fact substantial evidence that the spin degrees of  freedom in
different planes of a bilayer are strongly coupled.
Neutron scattering measurements have essentially only detected spin
fluctuations in which spin on adjacent $CuO_2$  layers are perfectly
anticorrelated \cite{Rossat-Mignon91,Shirane90,Tranquada92,Mook93}.
Moreover, the coupling between Cu spins on adjacent planes has been directly
measured in a recent NMR $T_2$ experiment, the Cu nuclear spins in one plane
of a bilayer were pumped and Cu nuclear spins in the other were measured
\cite{Stern95}.
This experiment determines the cross relaxation time $T_2^*$ which is given by
expression \cite{Monien95}
\begin{equation}
\frac{1}{T_2^*} =
\left [ \sum_q ( F_q \chi_{12}^{\prime} (q,0))^2 \right ]^{1/2}
\label{T_2^*}
\end{equation}
This expression is very similar to the expression (\ref{T_2}) for in-plane
relaxation rate, except that instead of a  single plane susceptibility
$\chi'(q,0)$ it contains $\chi'_{12} (q,0)$ which measures the
response of spins on plane 1 to the magnetic field on plane 2.

We assume that electrons on adjacent planes in bilayer interact via the
Hamiltonian
\begin{equation}
H_{\perp} = J_{\perp} \sum S_i^{(1)} S_i^{(2)}
\label{H_perp}
\end{equation}
If the interaction, $J_{\perp}$ is weak then the between the planes
susceptibility, $\chi_{12}$, may be calculated by perturbation theory
\cite{Millis95b} and is
\begin{equation}
\chi_{12}'(q,0) = J_{\perp} [\chi'(q,0)]^2
\label{chi_12}
\end{equation}
Experimentally the ratio $T_2/T_2^*$ grows from .14 at 200K to 0.28 at 100K.
This increase reflects the temperature dependence of $\chi^{\prime}$.
If $\chi^{\prime}$ is divergent at some wavevector $\bf Q$ as $T\rightarrow 0$
and is given by a scaling form $\chi^{\prime} = T^{- \alpha} f(|q-Q | T^{-x})$
(where $x$ and $\alpha$ are scaling exponents) then
from Eqs. (\ref{T_2},\ref{T_2^*},\ref{chi_12}) it may be shown that
$T_2 / T_2^* =c \; J_{\perp} \chi' (0,Q)$ with $c$ a constant of the
order of unity \cite{Millis95b}.
Thus, the observed maximal ratio of $T_2 / T_2^* =0.3$ implies that the
interplane coupling is not negligible, but still may be treated via
perturbation theory.
In this paper we shall show that the effect of this interplane
coupling on the fermions is large and in fact leads to the opening of a spin
gap.

Eq. (\ref{H_perp}) applies to $YBaCuO$ and to the multilayer $BSCCO$ compounds
in which the $Cu$ ion in one plane sits directly over the $Cu$ ion in the next
lower plane.
It does not apply to $La_{2-x}Sr_x Cu O_4$ compounds in which the crystal
structure is such that a $Cu$ ion on plane is coupled equally to four $Cu$ in
each adjacent plane, so Eq. (\ref{H_perp}) would become
\begin{equation}
H_{\perp'} = \sum_{a,ij} J^{\perp}_{ij}S_i^{(a)} S_i^{(a+1)}
\label{H_perp'}
\end{equation}
The crystal structure of $La_{2-x}Sr_x Cu O_4$ implies that $J_\perp(q)$
vanishes at $q=(\pi,\pi)$, so the enhancement of interplane pairing by
antiferromagnetic fluctuations is much less effective than in the $YBaCuO$ or
$BSCCO$ systems.

\section{MODEL}

\noindent
(i) {\it Single plane}.
In Section II we showed that experiment implies that a theoretical treatment
of spin gap effects in high-$T_c$ superconductors should involve pairing of
fermions in a spin liquid.
In this subsection we describe the model we use for the spin liquid in one
$CuO_2$ plane liquid and discuss different pairing mechanisms.

The low energy excitations of a spin liquid are $S=1/2$, charge 0 fermions,
$c^\dagger$, near a Fermi line, and a bosonic gauge field, $a$.
The action describing the spin liquid has been derived from more fundamental
models of correlated electrons by assuming that spin charge separation exists,
i.e. that the electron field $\psi^+$ may be written as the product of a
spinless, charge $e$ Bose field $b$ and a $S=1/2$ charge 0 Fermi field
$c^+$ as $ \psi^+ = c^+b$ and that the effect of the charge degrees of freedom
on the spin degrees of freedom is small.
These assumptions have been shown to be justified in the low doping, large
spin degeneracy limit of the $t-J$ model \cite{Ioffe89}.
Whether these assumptions are theoretically justifiable in the physically
relevant regime is still controversial.
We shall assume that they are because we see no other
way to explain the experimental data discussed in Section II.

The spin liquid model is specified by the action
\begin{eqnarray}
H=&&\sum_{p,\sigma}  c^{\dagger}_{p\sigma} \epsilon(p) c_{p\sigma} +
\sum_{p,k,\sigma}  c^{\dagger}_{p+k/2,\sigma} \vec{a}_k \vec{v}(p)
	c_{p-k/2,\sigma}
        \nonumber \\
+&& \sum_{p_i} W c^\dagger_{p_1,\alpha} c_{p_2,\alpha} c^\dagger_{p_3,\sigma}
        c_{p_4,\sigma} \delta(\sum p_i)
+ \frac{1}{4g_0^2} f_{\mu \nu}^2
\label{H}
\end{eqnarray}
Here, as usual, $f_{\mu \nu}= \partial_\mu a_\nu - \partial_\nu a_\mu$, $g_0$
is the bare fermion-gauge field interaction constant, ${\bf v}= \frac{\partial
\epsilon}{\partial {\bf p}}$, $W$ is a constant of short-range interaction
and $\sigma=1 \ldots N$ is a spin index.
In the physical case the spin degeneracy $N=2$.
The gauge field modifies the properties of the fermions.
These modifications have been studied in detail.
The results we shall need are:
(i) the electron self energy is \cite{Reizer89,Lee89}
$\Sigma(\epsilon) = \omega_0^{1/3} \epsilon^{2/3}$
(ii) the vertex, $\Gamma$, coupling the fermion spin to an external magnetic
field of wavevector $q$ becomes singular at $|q|=2p_F$ while at all other
wavevectors the vertex corrections are not singular \cite{Altshuler94}.
Specifically,
\begin{equation}
\Gamma = \Gamma_0 \left(\frac{\omega_0}{\Lambda}\right)^{\sigma}
\label{Gamma}
\end{equation}
Here $\Lambda$ is the largest of $v_F (|q|-2p_F)$, $\omega^{2/3}$, $T^{2/3}$
and
\begin{equation}
\omega_0= \left(\frac{1}{2\sqrt{3}} \right)^3
        \frac{v_F^3 g_0^4}{\pi^2 p_0}
\label{omega_0}
\end{equation}
is the upper cut-off scale determined  by the strength of the gauge field
fluctuations.
By ``$2p_F$'' we mean a wavevector $\bf Q$ which connects two points
on the Fermi  line with parallel tangents.
For a circular Fermi line any vector $\bf Q$ of magnitude $2p_F$ connects
two such points.
The exponent $\sigma$ has been calculated only in the limits
$N \gg 1$ and $N \ll 1$, by extrapolation these results to the physical value
$N=2$ we estimated \cite{Altshuler94} that $1 \gtrsim \sigma \gtrsim 1/3$.

The main effect of the self-energy renormalization is that the resulting
inelastic scattering rate $\sim T^{2/3}$ is so strong that no non-singular
interaction can lead to a BCS pairing of spinons \cite{Ubbens94a} (except via
a first order transition which is not observed);
therefore any spinon-based theory of the spin gap must involve a singular
interaction.

Two cases arise for the vertex renormalization.
If $\sigma < 1/3$, the spin physics is not modified in an essential way.
A $T=0$ critical point separates a $W < W_c$ phase with short range
spin correlations from a $W > W_c$ phase with long range order; the
appearance of the anomalous exponent $\sigma$ in the ``$2p_F$'' vertex
modifies the critical properties of the transition at $W = W_c$ as
discussed in detail in Ref. \cite{Altshuler95}.
However, if $\sigma > 1/3$ then the $0 \leq W < W_c$ phase is
anomalous, and exhibits a divergent ``$2p_F$'' spin susceptibility and power
law correlations
\begin{equation}
\chi(\omega , {\bf k} ) = \sqrt{\frac{\omega_0 p_0}{v_F^3}}
        \frac{1}{
        \left [ c_{\omega}
        \left ( \frac{|\omega|}{\omega_0} \right )^{2\sigma - 2/3}
        +  c_k \left(\frac{|k_{\parallel}| v_F}{\omega_0} \right)^{3\sigma-1}
        \right ] }
\label{chi}
\end{equation}
where $c_\omega,c_k \sim 1$.
Here we use local momentum coordinates associated with the Fermi line, namely
${\bf k}= {\bf Q} + {\bf e}_\parallel k_\parallel$ where $\bf Q$ connects
two points on the Fermi line with parallel tangents and ${\bf e}_\parallel$ is
the unit vector parallel to the Fermi velocity at these points.
These coordinates are generalization of radial $k_\parallel=|p|-2p_F$ and
angular coordinates for the case of non-circular Fermi line.
The experimental implications have been discussed elsewhere
\cite{Altshuler95}; in particular, it has been shown that the choice $\sigma
= 3/4$ yields rough agreement with experimental data.

Having discussed the spin liquid model we now consider possible pairing
interactions.
The short range interaction $W$ would lead to pairing in a Fermi liquid;
however, as mentioned above, for a spin liquid the inelastic scattering due to
the gauge field suppresses any second-order pairing instability due to a
non-singular interaction.
For this reason we believe the results obtained in Ref.\cite{Tanamoto92} do
not explain the spin gap phenomena.
Of course, a first order transition would still be possible \cite{Ubbens94a};
but there is no experimental evidence for this in high $T_c$ materials.
However, singular interactions exist.
One involves the gauge field, but this interaction is repulsive in all
channels for the model specified above and does not lead to pairing
\cite{Ioffe89}.
Another singular interaction comes from exchange
of long-ranged spin fluctuations;
these may arise either from proximity to a $T=0$ antiferromagnetic
transition or because $\sigma > 1/3$.

There are three types of antiferromagnetic transitions, distinguished by the
relation of the ordering wavevector, $\bf Q$, to $2p_F$.
If $|{\bf Q}| > 2p_F$, the fermion-fermion interaction mediated by spin
fluctuations is in fact not singular for the fermions near the Fermi line.
If $|{\bf Q}| = 2p_F$ and $\sigma < 1/3$, then the singularity of the
interaction is too weak to overcome the pairbreaking effect of the gauge field.
If $Q < 2p_F$, then one obtains a logarithmic divergence in the pairing
amplitude.
The theory of the pairing for $Q < 2p_F$ case may be derived by the following
the arguments given in Ref. \cite{Ioffe94} but replacing the factor
$\chi^2(\omega,q)$ by the first power $\chi(\omega,q)$ in Eq. (5) of Ref.
\cite{Ioffe94}.
The steps leading to Eq. (8) of Ref. \cite{Ioffe94} yield a logarithmic
divergence of the pairing kernel  in the gap equation.
However, we do not believe that the ${\bf Q} < 2p_F$ case is relevant to high
$T_c$ materials because the predicted temperature dependence of the copper NMR
$T_2$ rate is too weak and because neutron scattering has only observed
fluctuations peaked at wavevectors $Q \geq 2p_F$
\cite{Rossat-Mignon91,Shirane90,Mook93}.

If $\sigma > 1/3$, then the susceptibility is divergent at $Q \rightarrow
2p_F$, $(\omega, T) \rightarrow 0$.
However, as was shown in Ref\cite{Altshuler94} the same physics implies that
the fermion-fermion interaction is renormalized to zero (due to the
renormalization in the Cooper channel), so the divergence in
the susceptibility does not propagate into any other physical quantity.

\noindent
(ii) {\it Two Planes}:
The theoretical discussion in the previous subsection and the experimental
analysis of Section II implied that theories involving only a single $CuO_2$
plane could not explain the existence of a spin gap in a wide range of
temperatures above $T_c$.
In this subsection we extend the theory of the spin liquid to include
interplane coupling.
We assume that each plane is described by the Hamiltonian, Eq. (\ref{H}), and
that the only coupling between the planes is given by Eq. (\ref{H_perp}).
In particular, terms of the form $t_{\perp} c^{+(1)} c^{(2)} + h.c.$
are not allowed: there is no coherent hopping of spinons between planes.
The assumption that there is no interplane hopping of spinons may be
justified by extending the derivation \cite{Ioffe89} of the spin-liquid
action, $H$, to the two plane situation.
In the microscopic model there is a between-planes electron hopping
$t_{\perp}^{el}$ which, it is reasonable to assume, is much less than the
in-plane hopping, $t$.
It emerges from the theoretical derivation that one must have $t_{\perp}^{el}$
greater than  a critical value of order $t$ in order to have coherent
between-planes hopping of spinons.
If there is no coherent hopping then the leading coupling term is Eq.
(\ref{H_perp}).

We assume that $J_{\perp}$ is sufficiently weak that it may be treated in
perturbation theory.
This assumption is justified by the cross-relaxation
experiment of Stern et al \cite{Stern95}, as previously discussed.
The only effect we need consider the pairing interaction due to $J_{\perp}$.
If $Q < 2p_F$ then the relevant theory was given in \cite{Ioffe94,Ubbens94b}.
If $Q=2p_F$ and $\sigma < 1/3$ the theory is very similar.
As discussed above, we do not believe any of these starting points are
consistent with experiment.
We therefore study in Section IV the case $\sigma > 1/3$.
In Section V we treat the case of undamped spin fluctuations.

\section{GAP EQUATION; $\sigma > 1/3$}

In this section we consider pairing due to the between planes interaction, Eq.
(\ref{H_perp}).
In general, a pairing instability of a fermion system is signalled by a
divergence of the series of particle-particle ladder diagrams as shown in
Fig. 2. \cite{AGD}
In the present problem, the two lines correspond to fermions on different
planes and the between planes interaction $J_\perp$ is renormalized by the
gauge interaction.
This renormalization implies that the basic rung of the ladder is an effective
pairing interaction
\begin{equation}
V(\omega,k) = J_{\perp} \left[ \Gamma (\omega, k) \right]^2
\end{equation}
where $\Gamma$ is given in Eq. (\ref{Gamma}).
Note the absence in Fig. 2 of gauge field lines connecting fermions on one
plane to fermions on the other plane.
This absence follows from the assumption of no coherent fermion hopping
between the planes and is  the reason why the interaction is enhanced by the
gauge field, rather than suppressed by it as is the in-plane interaction.

\begin{figure}
\vspace{-2cm}
\centerline{\epsfxsize=6cm \epsfbox{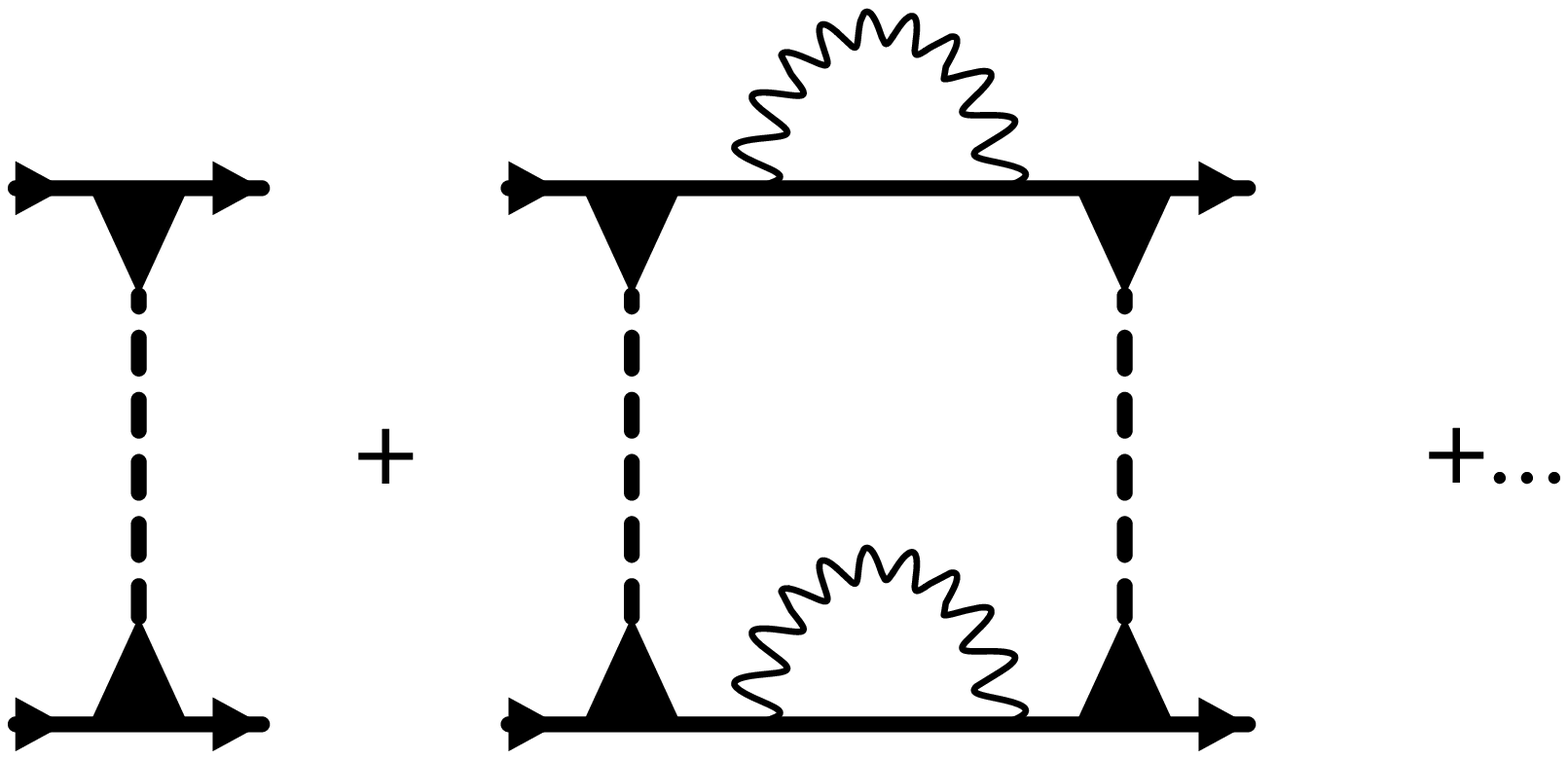}}
\vspace{-2cm}
{
FIG 2.
Ladder sum leading to pairing of fermions on adjacent planes. Solid lines
denote fermions propagators renormalized by the gauge field, dashed lines
denote spin-spin interaction, $J_\perp$, between the planes and solid
triangles denote vertices renormalized by the gauge field.
}
\end{figure}
\vspace{0.2in}

Because the interaction $V(\omega,k)$ connects fermions on different
planes it does not give rise to a fermion self energy in the leading order of
perturbation theory.
Diagrams of higher order in $J_\perp$ may be absorbed into the short-range
interaction $W$ between the  fermions on each plane and its renormalization by
gauge fields.
We have previously shown that $W$ is renormalized to zero by gauge fields, so
these diagrams may be neglected.

Therefore we may obtain the pairing effects  of the interaction $V$ from the
ladder sum in Fig. 2.
The analytic expression corresponding this diagram is, after integration over
$p_{\parallel}$
\begin{eqnarray}
\Delta ( p_{\perp}, \epsilon ) &=& \frac{T}{4 \pi}
\frac{J_{\perp} a^2 }{v_F}
\label{Delta} \\
\sum_{\omega} \int &&
	\frac{\Delta(p_{\perp}^{\prime}+p_\perp ,\epsilon + \omega)
	dp_{\perp}^{\prime}}
	{\left[ \left ( \frac{\omega}{\omega_0} \right )^{\sigma} +
	\left (\frac{v_F p_{\perp}^2}{p_0 \omega_0} \right )^{3\sigma/2}
	\right ]^2 ( \omega + \epsilon )^{2/3} \omega_0^{1/3}}
\nonumber
\end{eqnarray}
The integral on the right hand side of this equation is infrared dominated and
so may be evaluated by scaling (up to numerical factors).
We also use the definition of $\omega_0$, Eq. (\ref{omega_0}),
to eliminate the combination $\frac{\omega_0}{v_F}$
We find that $T_s$ is given by
\begin{equation}
T_s = \beta \omega_0 (J_{\perp} a^2 g^2 )^{\frac{3}{6 \sigma-2}}
\label{T_s}
\end{equation}
where $\beta$ is a numerical coefficient of the order of unity and $g^2$ is
renormalized fermion-gauge field interaction constant.
The value of $g^2$  has been estimated from the temperature dependence of the
resistivity \cite{Ioffe90}, and $\omega_0$ from the band structure.
Roughly, $1/(a^2 g^2) \sim 50 \; meV$ and $\omega_0 \sim 100 \; meV$.
$J_\perp$ can be estimated from the cross relaxation experiment to be $J_\perp
\sim 5 \; meV$.
In order for this $T_c$ to be of a reasonable order of magnitude, one must have
$6 \sigma \!-\!2 \approx 3$ i.e. $\sigma \approx 5/6$, because $J_{\perp}$ is
small.
This value of $\sigma$ would imply a $1/T_1 T \sim T^{-4/3}$, a somewhat more
rapid variation than is observed.

Although the divergent interaction exists at all points on the Fermi line, the
amplitude varies with position, $\theta$, because the gauge interaction
energy scale $\omega_0$ varies with $\theta$.
In fact, for the reasons discussed in Section VI of Ref. \cite{Altshuler94} we
expect this variation to be substantial in high $T_c$ materials.
Therefore we expect that as one lowers the temperature the gap
first appears at the points $\theta_{max}$ at which
$\omega_0$ is maximal, so $T_s= \beta \omega_0 (\theta_{max}) (J_{\perp}
a^2g^2)^{3/6 \sigma -2}$.
For $T$ near $T_s$ the gap function $\Delta ( \theta )$ will be very strongly
peaked about $\theta_{max}$; indeed from Eq. (\ref{Delta}) we see that
$\Delta ( \theta )$ decays away from $\theta_{max}$ as
$| \theta - \theta_{max} |^{-6 \sigma}$.
For $T=0$ the gap spreads over the whole Fermi surface but remains very
anisotropic:
\begin{equation}
\Delta(\theta) \approx \beta \omega_0(\theta)
(J_{\perp}a^2 g^2)^{\frac{3}{6\sigma-2}}
\label{Delta(theta)}
\end{equation}

Finally, we note that because the gap function is so strongly peaked at
particular points on the Fermi line, the energy is not very sensitive to the
symmetry of the gap.
In single-plane models, the pairing tends to be d-wave so that the two members
of the Cooper pair can avoid each other.
In the present model, the two members of the Cooper pair reside on different
planes, so d-wave pairing becomes favorable only when tunneling between
planes is included \cite{Kuboki95}.

\section{Gap Equation: Spin Waves}

In this section we consider the implications for the formation of the
spin gap of  an alternative picture of the origin of the antiferromagnetic
correlations.
We suppose that there are weakly damped propagating spin waves $\bf n$
with dispersion
$\omega^2=c^2 [ ( \vec{k} - \vec{Q} )^2 ] + \delta^2$ and we ask how these
lead to pairing.
Here $\vec{Q}= ( \pi , \pi )$, $c$ is the spin wave velocity and $\delta$ is
the spin wave gap.
In the ``quantum critical'' regime relevant for high $T_c$ superconductors,
$\delta = \alpha T$ with $\alpha \approx 1$.
We assume that $\Delta G = |Q|-2p_F >0$, so the low-energy spin waves lie
outside of the particle-hole continuum, and it is consistent to assume
they are coupled to the fermions, but are undamped.
The condition that the low-energy spin-waves are outside the particle-hole
continuum is (if $c \leq v_F$)
\begin{equation}
c \Delta G > T
\label{c_Delta}
\end{equation}
In high-$T_c$ materials we estimate from the Fermi line observed in Ref.
\cite{Shen95} that $\Delta G \gtrsim 0.1 \AA^{-1}$.
In insulating $La_2CuO_4$, $c = 0.75\; eV - \AA$ \cite{Hayden93}; the
previously discussed fits to $T_2/T_1 T$ in $YBa_2Cu_3O_{6.6}$ imply $c=0.35
\; eV - \AA$ \cite{Millis94}.
Adopting the latter value we find that the condition for the validity of this
assumption is $T \lesssim 0.035 \; eV \approx 400$K.
Thus the model may be adequate for discussion of
spin gap phenomena at $T \sim 150$K, but the relevance to the spin dynamics at
room temperature or above is questionable.

The action describing the coupling of fermions to spin waves is
\begin{eqnarray}
H_t    &=& H + H_{sw} \label{H_t} \\
H_{sw} &=& g \sum{\omega,\epsilon,k,p} \vec{n_{\omega,k}}
c^\dagger_{\epsilon+\omega,p+k}  \vec{\sigma} c_{\epsilon,p}
\nonumber \\
       &+& \sum_{\omega,k} \frac{1}{2D(\omega, k)} \vec{n}_{\omega,k}
		\vec{n}_{-\omega,-k} +
	J_\perp \vec{n}^{(1)}_{\omega,k} \vec{n}^{(2)}_{-\omega,-k}
\nonumber
\end{eqnarray}
Here $H$ is the fermion-gauge field action given in Eq. (\ref{H}), $\vec{n}$
describes the undamped spin fluctuation and the spin fluctuation propagator
$D(\omega,k)$ is
\begin{equation}
D(\omega, k)= \frac{u}{\omega^2 + (ck)^2 + \delta^2}
\label{D}
\end{equation}
and $u$ and $g$ are parameters with the dimension of energy.
As in the previous section, the interplane coupling leads to an effective
pairing interaction $V_{eff}$ given by
\begin{equation}
V_{eff}(\omega, k) = \frac{(gu)^2 J_{\perp}}{(\omega^2+(ck)^2+\delta^2)^2}
\label{V_eff}
\end{equation}
In order to account for the $Cu$ relaxation rate data, we must assume that for
$T>T_s$ the model is in the quantum critical regime in which $\delta=\alpha
T$.
Note that the momentum $k$ is measured from the commensurate antiferromagnetic
wavevector ${\bf Q}=(\pi,\pi)$ and that $V_{eff}$ is a very strongly peaked
function of $k$.
Thus, $V_{eff}$ scatters a fermion of momentum $\bf p$ mostly to states with
momentum very near $\bf p+Q$ and if Eq. (\ref{c_Delta}) is satisfied and $\bf
p$ is on the Fermi surface, these final states are far from the Fermi surface.
This strong momentum dependence means that the gap equation decomposes into
two equations, one expressing the gap at momentum near $\bf p$,
$\Delta_\epsilon(p)$, in terms of the Green function at momenta near $\bf p+Q$
and another one expressing the gap at momentum $\bf p+Q$,
$\Delta^{*}_\epsilon(p)$, in terms of the Green function at momenta near $p$.
Specifically,
\begin{eqnarray}
\Delta_\epsilon(p) &=& \sum_{\omega,q} \frac{ T V_{eff}(\omega,q)
	\Delta_{\epsilon+\omega}^{*}(p+q)}
	{\omega_0^{2/3} |\epsilon \!+\! \omega|^{4/3} \!+\!
	\tilde{\zeta}_{p+q}^2 \!+\!
	[\Delta_{\epsilon+\omega}^{*}(p+q)]^2}
\label{Delta_1}
\\
\Delta_\epsilon^{*}(p) &=& \sum_{\omega,q} \frac{ T V_{eff}(\omega,q)
	\Delta_{\epsilon+\omega}(p+q)}
	{\omega_0^{2/3} |\epsilon \!+\! \omega|^{4/3} \!+\! \zeta_{p+q}^2
	\!+\! [\Delta_{\epsilon+\omega}(p+q)]^2}
\label{Delta_2}
\end{eqnarray}
Here $\zeta_p$ is the fermion energy, $\tilde{\zeta_p}=\zeta_{p+Q}$; for
circular Fermi line  $\zeta_p=v_F(|{\bf p}|-p_F)$.
Because $V_{eff}$ is a strongly peaked function of $q$, we may neglect the $q$
in denominator of (\ref{Delta_1}).
The magnitude of the denominator is then controlled by the magnitude of
$\zeta_{p+Q}$.
This depends on the position of $p$ along the Fermi line.
The minimal value of $\zeta_{p+Q}$ is $v_F \Delta G$ and the maximal value is
of the order of $\epsilon_F$.
Because $\int V_{eff}(\omega,q) d^2 q$ is a singular function of $\omega$, the
frequency sum is dominated by $\omega \sim {\mbox max}(T,\delta)$ and we
expect that $\delta \sim T$, so by (\ref{c_Delta}) we may also neglect the
frequency and $\Delta^{*}$ dependence of the fermionic denominator in
(\ref{Delta_1}).
The first equation (\ref{Delta_1}) then becomes a linear convolution equation;
it can be combined with the second equation (\ref{Delta_1}) giving an
equation which contains only the gap function in the vicinity of the Fermi line
\begin{equation}
\Delta_\epsilon(p) = \frac{T}{\zeta_{p+Q}^2}
	\sum_{\omega,q} \frac{ W_{eff}(\omega,q)
	\Delta_{\epsilon+\omega}(p+q)}
	{\omega_0^{2/3} |\epsilon+\omega|^{4/3} + \zeta_{p+q}^2 +
	\Delta_{\epsilon+\omega}^2(p+q)}
\label{Delta_eq}
\end{equation}
with the singular kernel
\begin{equation}
W_{eff}(\omega,q) = T \sum_{\eta,k} V_{eff}(\eta,k) V_{eff}(\omega-\eta,q-k)
\label{W_eff_1}
\end{equation}
In the gap equation (\ref{Delta_eq}) the assumption that $p$ is on the Fermi
surface means that the singularities in the fermion denominator must be also
considered and compared to the singularities of the kernel $W_{eff}$.
The fermion propagator depends sensitively only on the component of momentum
normal to the Fermi surface and, as we show below,  $\Delta_\epsilon(p)$ also
changes smoothly along the Fermi surface.
We may therefore integrate over the component along the Fermi line, $q_\perp$,
immediately, obtaining an equation similar to (\ref{Delta_eq}) but with
$q_\perp=0$ and a modified kernel
\begin{eqnarray}
W_{eff}(\omega,q_\parallel) &=& \frac{T(gu)^4 J_\perp^2}{16c^2}
\sum_{\eta,k_\parallel} \frac{1}{[\eta^2+(ck_\parallel)^2+\delta^2]^{3/2}}
\nonumber \\
&& \frac{1}{[(\omega-\eta)^2+(ck_\parallel-cq_\parallel)^2+\delta^2]^{3/2}}
\label{W_eff_2}
\end{eqnarray}
Two cases then arise. If
\begin{equation}
\frac{\delta}{c} < \frac{{\mbox max}(\omega_0^{1/3} T^{2/3}, \Delta)}{v_F}
\label{delta_c}
\end{equation}
then the singularity in $W_{eff}$ is dominant and we may perform the
$k_\parallel$ and $q_\parallel$ integrals obtaining
\begin{equation}
\Delta_\epsilon(p) = \frac{T}{\zeta_{p+Q}^2}
	\sum_{\omega} \frac{ W_{eff}(\omega)
	\Delta_{\epsilon+\omega}(p)}
	{\omega_0^{2/3} |\epsilon+\omega|^{4/3} + \zeta_{p}^2 +
	\Delta_{\epsilon+\omega}^2(p)}
\label{Delta_eq_1}
\end{equation}
with
\begin{eqnarray}
W_{eff}(\omega) = \frac{T(gu)^4 J_\perp^2}{16 \pi^2 c^4}
\sum_{\eta} \frac{1}{[\eta^2+\delta^2]}
\frac{1}{[(\omega-\eta)^2+\delta^2]}
\label{W_eff_3}
\end{eqnarray}
However, if the condition (\ref{delta_c}) is not satisfied, then the
singularity in the fermion denominator is dominant and we obtain
\begin{equation}
\Delta_\epsilon(p) = \frac{T}{\zeta_{p+Q}^2}
	\sum_{\omega} \frac{ W_{eff}(\omega)
	\Delta_{\epsilon+\omega}(p)}
	{\sqrt{\omega_0^{2/3} |\epsilon+\omega|^{4/3} +
	\Delta_{\epsilon+\omega}^2(p)}}
\label{Delta_eq_2}
\end{equation}
with
\begin{eqnarray}
W_{eff}(\omega) &=& \frac{T(gu)^4 J_\perp^2}{32 \pi  c^3 v_F}
\sum_{\eta} \int \frac{dk}{[\eta^2+\delta^2+k^2]^{3/2}}
\nonumber \\
&& \frac{1}{[(\omega-\eta)^2+\delta^2+k^2]^{3/2}}
\label{W_eff_4}
\end{eqnarray}
This system of equations may be solved to determine $T_s$ and $\Delta(T)$.
At low $T$, or for not too small $c$ we expect Eq. (\ref{delta_c}) to be
satisfied, so we shall consider in detail only the
Eqs.(\ref{Delta_eq_1}-\ref{W_eff_3}).
The other case leads to very similar results.

To estimate $T_s$ we linearize Eq. (\ref{Delta_eq_1}) and, because the sums
are infrared dominated take only the contribution from the lowest Matsubara
frequency.
We find
\begin{equation}
T_s(p_\theta)=\left[ \frac{J_{\perp} (gu)^2}
	{4 \pi^{5/3} c^2 \zeta_{p_\theta+Q} \omega_0^{1/3}
\alpha^2} \right]^{3/5}
\end{equation}

Similarly, if the condition (\ref{delta_c}) is not satisfied we find from
(\ref{Delta_eq_2}) and (\ref{W_eff_4}) the slightly different formula
\begin{equation}
T_s(p_\theta)=\left[\sqrt{3} \frac{J_{\perp} (gu)^2}
	{16 \pi^{1/3} c^{3/2} v_F^{1/2} \zeta_{p_\theta+Q} \omega_0^{1/6}
\alpha^{5/2} } \right]^{6/11}
\end{equation}

As found in the previous section, the onset of the spin gap is angle
dependent.
It appears first at the angle, $\theta^{*}$,  for which $\zeta_{\bf p+Q}$ has
a minimum, and as $T$ is lowered spreads over the Fermi line.
Because the interaction is more singular than in the case of the damped spin
waves considered in the previous section, $T_s$ goes as a smaller power of
$J_\perp$.
Roughly, $\Delta$ is large for $\theta_p$ such that $T_s(p_\theta)>T$, and
drops rapidly for larger $\theta$.
We denote the interval in which $\Delta$ is large by
$(\theta^*+\theta_0,\theta^*-\theta_0)$.
We note that
\begin{equation}
\zeta_{p_\theta+Q}=\zeta_* + \epsilon_0(\theta-\theta^*)^2
\end{equation}
where $\epsilon_0$ is an energy scale of the order of $\epsilon_F$.
Thus, for $T$ very near $T_s$, $\theta_0$ is given by
\begin{equation}
\theta_0 \sim \left(\frac{T_s-T}{T_s}\right)^{1/2}
	\left(\frac{\zeta_*}{\epsilon_0}\right)^{1/2}
\end{equation}
For $T$ much less than $T_s$
\begin{equation}
\theta_0 \sim \left(\frac{T_c}{T}\right)^{5/6}
	\left(\frac{\zeta_*}{\epsilon_0}\right)^{1/2}
\end{equation}
For $|\theta-\theta^*|>\theta_0$ $\Delta(\theta)$ is induced by the value of
$\Delta$ inside the interval $(\theta^*+\theta_0,\theta^*-\theta_0)$.
{}From Eqs. (\ref{Delta_eq_1},\ref{W_eff_3}) it can be seen that the kernel is
so sharply peaked that
\begin{equation}
\Delta(\theta) \sim \Delta(\theta_0) \frac{\theta_0^4}{(\theta-\theta_0)^4}
\end{equation}
Finally, at $T=0$ one finds
\begin{equation}
\Delta(\theta) \sim \left[ \frac{J_\perp (gu)^2}{4 \pi c^2 \zeta_{p+Q}
\omega_0^{1/3} \alpha^2} \right]^{3/5}
\label{Delta_3}
\end{equation}

In words, because the interaction is so strongly peaked, the value of the
$T=0$ gap is controlled by the energy, $\zeta_{p+Q}$, of the intermediate
state of momentum $p+Q$.

\section{Conclusion}

We have argued that any theory of the spin gap observed in $YBa_2Cu_3O_{6+x}$
or $YBa_2Cu_4O_8$ must be based on a pairing instability of a Fermi sea of
chargeless fermions (i.e. a spin liquid) in the presence of antiferromagnetic
correlations.
The large inelastic scattering found in the models of spin liquids leads to
strong pairbreaking effects which can only be overcome by singular pairing
interaction.
The most plausible origin of this interaction is a between planes coupling
enhanced by in-plane antiferromagnetic correlations.
We considered two specific models: a spin liquid with $2p_F$ over damped
magnetic correlations induced by a gauge field interaction and a spin liquid
coexisting with weakly damped antiferromagnetic spin waves.
(A third possible model, namely a spin liquid with magnetic correlations
induced by tuning a four fermion interaction, has been considered elsewhere
\cite{Ioffe94,Ubbens94a})
The former model is clearly appropriate to optimally doped materials, where
there is convincing photoemission evidence for a large Fermi line.
We expect by continuity that it is also appropriate for dopings somewhat below
optimal.
Weakly damped spin waves exist in the insulating ``parent compaund''
materials, and it has been proposed that low doping induces a gap (but no
damping) in the spin wave spectrum as well as a particle hole continuum of
spin excitations.
This picture might be justified if there is a small Fermi line (i.e. ``hole
pockets''), but if there is a large (Luttinger) Fermi line the absence of spin
wave damping is difficult to justify.
It is controversial which picture applies to the most extensively studied spin
gap compounds $YBa_2Cu_3O_{6.7}$ or $YBa_2Cu_4O_8$.
We analysed both models, obtaining estimates for the onset temperature, $T_s$,
of the spin gap and the angular dependence of the gap function.

In both models $T_s$ scales as a power of the between planes coupling
$J_\perp$, and the gap function is sharply peaked about particular regions of
the Fermi line.
In both models, pairing of spinons does not imply a true thermoduynamic
transition \cite{Ioffe89}.
$T_s$ is a crossover temperature and superconductivuty sets in at a lower
temperature $T_c$ at which the charge carrying bosons condense.

In the overdamped case the physics is controlled by divergences in the
fermion $2p_F$ response function due to the singular gauge field interaction.
These divergencies are characterized by an exponent $\sigma$ which depends
only on the fermion spin degeneracy $N$ and by an energy scale $\omega_0$
defined in Eq. (\ref{omega_0}) which varies substantially as one moves along
the Fermi line.
We found that $T_s \propto \omega_0^{max} J_\perp^\frac{3}{6\sigma-2}$
($\omega_0^{max}$ is the maximal value of $\omega_0(\theta)$ on the Fermi
line which occurs at $\theta^*$).
At $T\approx T_s$ the gap is very sharply peaked at $\theta=\theta^*$.
As $T$ is lowered the region where gap is appreciable grows and the $T=0$ gap
$\Delta(\theta) \propto \omega_0(\theta) J_\perp^\frac{3}{6\sigma-2}$.
Evidently the result depends sensitively on the exponent $\sigma$ which has
been estimated to be in the range $1 \gtrsim \sigma \gtrsim 1/3$.
Consistency with NMR for $T \gtrsim T_s$ requires that
$1 \gtrsim \sigma \gtrsim 2/3$; for this range of $\sigma$ $T_s$ is
proportional to $J_\perp$ to a power of order 1 and is of the correct order of
magnitude, but the precise value depends on numerical coefficients which are
not known.

For $T_s \gtrsim T \gtrsim 0$, the spin gap is appreciable over a part of the
Fermi line, and suppresses the contribution of that part of the Fermi line to
the uniform susceptibility and NMR relaxation rates.
We see from Eqs. (\ref{omega_0},\ref{Delta(theta)}) that the gap is largest
along the zone diagonal (where $v_F$ is largest) and smallest at the zone
corners where $v_F$ becomes very small.
All parts of the Fermi line make roughly equal contribution to the oxygen
relaxation rate and uniform susceptibility (although the logarithmic
divergence associated with the van Hove singularity may emphasize the corners
to some extent), so we may roughly estimate the suppression of these
quantities from the fraction of the Fermi line which is gapped.
The copper relaxation rate is more complicated.
It is dominated by the $2p_F$ fluctuations which lead to
\[
\frac{1}{T_1T} \sim \frac{\omega_0^{2\sigma-2/3}}{p_0^{1/3}}
\]

Thus the contribution to the $Cu$ relaxation rate is largest where $\omega_0$
is largest (i.e. along the zone diagonal) so one would expect that in this
model the formation of the spin gap would suppress the $Cu$ relaxation rate
more strongly than the oxygen rate.
However, the contributions of the ungapped portions of the Fermi line continue
to grow as $T$ is lowered, so the maximum in the $Cu$ $1/T_1T$ relaxation rate
occurs at a $T<T_s$ determined by the interplay between these two effects.

We now consider the underdamped case.
The basic assumption is that there are two distinct types of spin excitations:
propagating antiferromagnetic spin waves and particle-hole continuum of spinon
excitations.
This picture has been derived \cite{Chubukov93} from a microscopic Hamiltonian
using the assumption that the doping is so low that long range magnetic order
is present, and it is plausible that it may apply to lightly doped high $T_c$
materials which lack long range order if these materials do not have large
(Luttinger) Fermi line but have instead ``hole pockets''.
An advantage of this picture is that large $q$ properties are dominated by
spin waves, which explain in a natural way the strong $T$ dependence of the
$Cu$ $1/T_1T$ and $1/T_2$ rates observed experimentally.
The disadvantage of this picture is that the same spin waves would give a
factor of six too large contribution to $d\chi/dT$ at high temperatures so to
account for the observed $d\chi/dT$ one must assume that the particle-hole
continuum leads to a large negative contribution to $d\chi/dT$ which almost
precisely cancels the spin wave contribution.

We studied the pairing of fermions in the presence of the undamped spin waves.
The pairing interaction is very strongly peaked at the antiferromagnetic
wavevector ${\bf Q}=(\pi,\pi)$, so the dominant process in the gap equation is
a virtual scattering from a state $\bf p$ on a Fermi line to one at $\bf p+Q$
away from the Fermi line.
We found that the resulting pairing interaction is very strong, the onset
temperature $T_s \propto J_\perp^{3/5}$.
As in the overdamped case, the gap has a strong angular dependence, it appears
first at a particular point on the Fermi line and spreads over it as $T$ is
decreased.
However in the underdamped case the angular dependence of the gap is
controlled by the energy of the intermediate state, $\zeta_{p+Q}$.
The pairing affects the particle-hole contribution to physical response
functions but does not directly affect the spin wave contribution.
Thus, the copper relaxation rate is not significantly affected by the pairing,
but the oxygen relaxation rate and uniform susceptibility are.
The detailed temperature dependence is determined by the way in which the gap
spreads over the Fermi line as the temperature is decreased, and this depends
sensitively on the shape of the Fermi line and, in particular, on its
curvature.

This picture provides a natural explanation of the magnetic properties of
lightly doped $YBaCuO$ materials ($YBa_2Cu_3O_{6+x}$ with $0.45<x<0.6$) in
which the $T_s$ inferred from the uniform susceptibility is of the order
$300K$ \cite{Alloul89} while strong antiferromagnetic fluctuations with no
evident gap have been observed in neutron scattering \cite{Tranquada92}.
However, in this picture the coincidence of $T_s$ with the maximum of the $Cu$
$1/T_1T$ rate observed in $YBa_2Cu_3O_{6.7}$ and $YBa_2Cu_4O_8$ is an
accident.

Thus, we believe that the overdamped picture is more appropriate to materials
with doping greater than one in $YBa_2Cu_3O_{6.7}$, while the underdamped
picture may be more appropriate for more lightly doped materials.
This general point of view has been advocated by many authors \cite{Lee90} and
has been emphasized recently in Ref. \cite{Sokol93,Chubukov93}.

\end{multicols}

\end{document}